\documentclass[reprint,amsmath,amssymb,aps,showkeys]{revtex4-2}

\usepackage{slashed}
\usepackage{amsmath}

\usepackage{graphicx}
\usepackage{caption}
\usepackage{subcaption}
\usepackage[justification=raggedright,format=hang]{caption}
\usepackage{dcolumn}
\usepackage{bm}
\usepackage{hyperref}
\usepackage[mathlines]{lineno}
\usepackage{xcolor}
\usepackage[makeroom]{cancel}
\usepackage{upgreek}

\usepackage{graphicx}
\usepackage{dcolumn}
\usepackage{bm}
\usepackage{bm,latexsym,amsmath,amssymb,amsfonts,color}

\begin{document}

\preprint{APS/123-QED}

\title{\textbf{Exact black holes and black branes with bumpy horizons \\ supported by superfluid pions} 
}%
\author{Fabrizio Canfora}
\email{fabrizio.canfora@uss.cl}
\affiliation{Centro de Estudios Cient\'{\i}ficos (CECS), Casilla 1469, Valdivia, Chile.}
\affiliation{Facultad de Ingeniería y Tecnología, Universidad San Sebastián, General Lagos 1163, Valdivia 5110693, Chile.}
\author{Andr\'es Gomberoff}
\email{andres.gomberoff@umayor.cl}
\affiliation{Facultad de Ciencias, Ingeniería y Tecnología, Universidad Mayor, Avenida Manuel Montt 367, Santiago, Chile}
\affiliation{Centro Multidisciplinario de F\'isica, Vicerrector\'ia de Investigaci\'on, Universidad Mayor, Camino La Pir\'amide 5750, Santiago, Chile.}
\author{Carla Henr\'iquez-Baez}
\email{carla.henriquez@umayor.cl}
\affiliation{Centro Multidisciplinario de F\'isica, Vicerrector\'ia de Investigaci\'on, Universidad Mayor, Camino La Pir\'amide 5750, Santiago, Chile.}
\author{Aldo Vera}
\email{aldo.vera@umayor.cl}
\affiliation{N\'ucleo de Matem\'atica, F\'isica y Estad\'istica, Universidad Mayor, Avenida Manuel Montt 367, Santiago, Chile}
\affiliation{Centro Multidisciplinario de F\'isica, Vicerrector\'ia de Investigaci\'on, Universidad Mayor, Camino La Pir\'amide 5750, Santiago, Chile.}

\begin{abstract}
We present exact solutions of the Einstein--$SU(2)$ non-linear sigma model in $3+1$ spacetime dimensions, describing bumpy black holes and black branes. Using an Ansatz for superfluid pion multi-vortices, the matter sector reduces to a first-order BPS system, while the Einstein equations reduce to a Liouville equation with a smooth source governing the horizon deformation.
These solutions describe horizons of different constant curvatures, with nontrivial bumpy geometries protected by an integer topological invariant, namely the vorticity, which also controls the number of bumps and the black hole thermodynamics.
Remarkably, such horizons arise in a minimal and physically motivated matter model, without invoking exotic fields or modified gravity. The physical implications of these results in holography and astrophysics are briefly described.
\end{abstract}

                        
\maketitle

\section{Introduction}

Black holes and black branes are fundamental objects in both theoretical physics and astrophysics, whose properties are intrinsically determined by the shape of their horizons. In General Relativity (GR), in $3+1$ dimensions without cosmological constant, there are many rigidity theorems which restrict both the topology and the curvature of the horizon \cite{Israel:1967wq, Carter:1971zc,Hawking:1971vc,Robinson:1975bv,Friedman:1993ty}. Now, the presence of a negative cosmological constant allows event horizons of constant curvature with toroidal or hyperbolic shapes \cite{Gibbons:1977mu,Linet:1986sr,Lemos:1994xp,Birmingham:1998nr,Cai:1996eg,Vanzo:1997gw,Klemm:2014rda}, and in space-time dimensions higher than four, horizons with exotic topologies (such as black ring and black Saturn) can appear \cite{Emparan:2001wn,Elvang:2007rd} (see also Refs. \cite{Gubser:2001ac,Licht:2020odx,Iguchi:2007is}).

Black objects with bumpy horizons are particularly relevant because they appear as transitional states in black hole mergers, as well as in the black string instability process \cite{Gregory:1993vy,Lehner:2010pn,Figueras:2015hkb}. Also, significant variations in the measurement of the black hole shadow and its emission of gravitational waves appear, as was shown in Ref. \cite{Collins:2004ex}, where the authors studied the gravitational effects of a hypothetical bumpy black hole based on deformations of the Kerr metric (see Refs. \cite{Herdeiro:2014goa,Herdeiro:2016tmi,Doneva:2017bvd,Silva:2017uqg,Doneva:2022ewd} for the construction of black holes and compact objects that exhibit geometric deviations from the classical solutions of GR by coupling scalar and vector fields).

Nevertheless, black objects with non-constant local curvature, such as black bottles \cite{Chen:2016rjt}, spindle black holes \cite{Ferrero:2020twa,Ferrero:2020laf,Giri:2021xta,Boido:2022mbe}, or bumpy black holes \cite{Novikov:1992gow,Collins:2004ex,Johannsen:2011dh,Vigeland:2011ji,Emparan:2014pra,Dubovsky:2007zi}, usually require additional ingredients to circumvent the theorems of GR, such as the coupling with exotic matter fields to obtain a hair that can locally deform the event horizon \cite{Hristov:2023rel,Suh:2023xse,Farhangkhah:2014zka}. Even with the addition of exotic ingredients, the stability of these types of configurations is not guaranteed.

A fundamental question therefore remains open: can bumpy horizons be supported in \(3+1\) dimensions by a realistic matter field? This question is extremely relevant, for instance, in the analysis of the exterior gravitational field of compact objects with superfluid vortices piercing the external surface from the inside. The evidence presented here confirms this to be the case.  We consider GR with a cosmological constant coupled to a non-linear sigma model (NLSM) describing the low-energy dynamics of pions. Using an Ansatz for superfluid pion multi-vortices \cite{Canfora:2024mkp}, the gravitational back reaction reduces to a Liouville equation with a smooth source that can be solved analytically. The matter content represents the pions superfluid, with the phase postulated by Feynman and Onsager \cite{onsager1949,feynman1955}. The resulting horizons are protected by an integer topological charge, the vorticity, which stabilizes the bumps against gravitational stretching.

From the holographic viewpoint, bumpy black branes provide a natural realization of translational symmetry breaking in hydrodynamics and offer a concrete realization of the mechanisms underlying entropy production and viscosity bounds \cite{Hartnoll:2016tri,Kovtun:2004de}.

\section{The Einstein-NLSM}

The low-energy dynamics of pions is described at leading order by the $SU(2)$ NLSM, which we adopt herein. In $D=4$ space-time dimensions and in the presence of a cosmological constant, the Einstein $SU(2)$-NLSM action is given by
\begin{gather}
I[g,U]=\frac{1}{2\kappa}\int_{\mathcal{M}} d^{4}x\sqrt{-g}\left(R-2\Lambda+
\frac{K}{4}\left(\mathrm{Tr}[L^{\mu }L_{\mu }] \right)\right) \  ,  \label{I}  \\
 L_{\mu}=U^{-1} \nabla_{\mu} U = L^{i}_{\mu}t_{i} \ , \quad t_{i}=i\sigma_{i} \ , \notag
\end{gather}
where $R$ is the Ricci scalar, $g$ is the metric determinant, $U\left(x \right) \in SU(2)$ is the pionic field and $\sigma_i$ are the Pauli matrices. In Eq. \eqref{I}, $\kappa=8 \pi G$ is the gravitational constant, $\Lambda$ is the cosmological constant and $K$ is a dimensionless coupling constant defined in terms of the pion decay constant  as $K=\frac{\kappa f_\pi^2}{2}$.\footnote{The pion mass can be safely neglected in the present context as it is orders of magnitude smaller than the solitons of the theory.}

The pionic field in the exponential representation is written as 
\begin{gather}
U=\cos (\alpha ) \, \boldsymbol{1}_2+\sin (\alpha )n_{i}t^{i}\ ,  \label{U} \\
n_{i}=\{\sin \Theta \cos \Phi ,\sin \Theta \sin \Phi ,\cos \Theta \}\ , 
\notag
\end{gather}%
where $\alpha =\alpha (x^{\mu })$, $\Theta =\Theta (x^{\mu })$, $\Phi
=\Phi(x^{\mu })$ are the three degrees of freedom of the $U$ field, and $%
\boldsymbol{1}_2$ denotes the $2\times2$ identity matrix.

We take the following Ansatz for the metric 
\begin{eqnarray} \label{metric}
    ds^2=-f(r)dt^2+\frac{dr^2}{f(r)}+r^2 e^{P(x, y)}\left(dx^2 +dy^2 \right) \ ,
\end{eqnarray}
where $(x,y)$ represent the coordinates on the transverse section, providing the base space for the pionic vortex configuration and the resulting horizon deformations discussed below.

For the matter fields we use the Ansatz introduced in Ref. \cite{Canfora:2024mkp}, which is suitable to find the gravitational field corresponding to BPS multi-vortices: 
\begin{gather} \label{matter}
    \alpha=\alpha(x, y) \ , \qquad \Theta= \frac{\pi}{2} \ , \qquad \Phi=\Phi(x, y) \ .
\end{gather}
While our Ansatz may appear restrictive, it describes a configuration characterized by a single Killing vector associated with time translations; consequently, these solutions lack both spherical and axial symmetry. This reduction in symmetry is a direct consequence of the pionic vortices distributed at the horizon, where these topological solitons break rotational invariance and induce the characteristic deformations (bumps) of the geometry. Furthermore, as we will show below, despite that the matter field does not depend on the radial coordinate, the configuration possesses a finite mass and remains topologically protected.

The fact that $\partial_t$ is a Killing vector implies the conservation of the current $j^\mu=\sqrt{-g} \, T^\mu_0$. However, $j^r=0$, and therefore the integral of the energy density of the pionic field ($\varepsilon =j^0=\sqrt{-g} \, T^{t}_{\, \, t}$) over surfaces of constant $r$ is conserved.

The energy density is given by
\begin{eqnarray*}
    \varepsilon=  \frac{K}{2\kappa}  \left\{(\partial_{x} \alpha)^2 +(\partial_{y}\alpha)^2 + \sin^2 \alpha\left( (\partial_{x}\Phi)^2 +(\partial_{y}\Phi)^2 \right)\right\}  \ ,  
\end{eqnarray*}
and then, the energy may be written in a BPS form as
\begin{eqnarray}\label{energy}
   E &=&  \frac{K}{2\kappa}   \int d^2 x\left\{ (\partial_x\alpha \mp \sin\alpha\,\partial_y\Phi)^2
+(\partial_y\alpha \pm \sin\alpha\,\partial_x\Phi)^2 \right\}\nonumber \\ && \pm  \frac{K}{\kappa} \int d^2x \sin\alpha\,
(\partial_x\alpha\,\partial_y\Phi-\partial_y\alpha\,\partial_x\Phi) \ .
\end{eqnarray}
The last integral is a topological term, proportional to
\begin{equation}
Q
=
\frac{1}{4\pi}\int_{\Sigma}\sin\alpha\; d\alpha\wedge d\Phi \ .
\label{eq:EBPS}
\end{equation}
This topological charge represents the vorticity of the superfluid pions as $\Phi$ represents the phase of the superfluid. Since the fields $(\alpha,\Phi)$ parametrize the coset $SU(2)/U(1)\simeq S^2$, it measures the homotopy invariant that counts how many times $\Sigma$ wraps around $S^2$. In the next section we compute it explicitly.

Thus, Eq. \eqref{energy} implies that $E\geq 4\pi|Q|K/\kappa$, which is saturated in the BPS case when the squared terms in the first line of  Eq. \eqref{energy} vanish. It turns out that the BPS condition ensures that the equations for the matter field obtained by varying $U$ in the action \eqref{I} are satisfied. Using the  \emph{lower signs} in Eq. \eqref{energy}, the BPS equations are
\begin{equation}
\partial_x\alpha=-\sin\alpha\,\partial_y\Phi \ ,
\qquad
\partial_y\alpha=\sin\alpha\,\partial_x\Phi \ .
\label{BPS}
\end{equation}

The opposite choice of signs corresponds to reversing the orientation of the
map $(\alpha,\Phi):\Sigma\to S^2$ and simply flips the sign of the topological
term.

Despite the non-trivial dependence of the metric both on the $(x,y)$ coordinates of the horizon and on the radial coordinate $r$, the full system of Einstein-NLSM decouples in a $(x,y)$-sector and a $r$-sector. This is why we can find fully analytic solutions of black holes and black branes with just one Killing field.

It is a direct computation to show that all the solutions of the above BPS system are also solutions of the second order field equations:
\begin{align}
    \Delta_F \Phi + 2 \cot(\alpha) (\vec\nabla \alpha \cdot \vec\nabla \Phi) =&  0 \ , \\
    \Delta_F \alpha - \frac{1}{2} \sin(2\alpha) (\vec\nabla \Phi)^2 =&  0 \ ,
\end{align}
being $\Delta_F= \partial_x^2+\partial_y^2$ the flat-space Laplacian. Note that in Eqs. \eqref{BPS} the metric does not appear, so that the above equations are exactly the same as in the flat space-time case \cite{Canfora:2024mkp}.

\section{Black holes and black branes}

It is convenient to introduce a new function, $H(x, y)$, defined by
\begin{equation}
    \alpha= 2 \arctan(e^{H}) \ \label{alpha} \ .
\end{equation}
In terms of it, the BPS system in Eq. \eqref{BPS} become linear, 
\begin{equation} \label{BPSa}
    \partial_x H + \partial_y \Phi = 0 \ ,\ \ \  
    \partial_y H - \partial_x \Phi = 0 \ .
\end{equation}
These equations imply that, on the complex plane $z=x+iy$, the combination
\begin{equation}
g(z)= H(z)-i\Phi(z) \ , 
\end{equation}
is a holomorphic function. For an elementary solution of the matter fields, it is
natural to look for circularly symmetric configurations around the origin.
However, the only holomorphic function whose real and imaginary parts are both
independent of the polar angle $\phi$ in $z=\rho e^{i\phi}$ is a constant. 
Therefore, exact circular symmetry is incompatible with holomorphy. The simplest nontrivial possibility is to require only the real part of $H$ to be
radially symmetric. This selects
\begin{equation}\label{complex}
g(z)= q\log z \ ,
\end{equation}
being $q$ a constant, for which $H=q\log\rho$, while $\Phi=-q\phi$. Since the equations are linear, more
general solutions can be constructed by superposition, leading to the multi-center
configuration
\begin{gather}
H(\vec{\rho})=\sum_i q_i \log\left|\vec{\rho}-\vec{\rho}_i\right| \ ,
\label{Hmulti}\\
\Phi(\vec{\rho})=-\sum_i q_i \arctan\!\left(\frac{y-y_i}{x-x_i}\right) \ , \label{Phimulti}
\end{gather}
with $\vec{\rho}=(x,y)$. We see that the above solution has the same form of the superfluid quantum phase postulated by Feynman and Onsager \cite{onsager1949,feynman1955,donnelly1991,fetter2009}). 

Note, from Eq. \eqref{Hmulti}, that the sum of the vorticities $q_i$ on a compact surface $\Sigma$ should vanish, because
\begin{equation}\label{qz}
    \sum_i q_i= \frac{1}{2\pi}\int_\Sigma \sqrt{g_\Sigma} \,\Delta H = 0 \ .
\end{equation}
On the other hand, the Einstein equations are reduced to two equations for its components in Eq. \eqref{metric}. The first one determines the lapse function, obtaining 
\begin{equation}
    f(r)=\gamma -\frac{2m}{r}-\frac{\Lambda }{3}r^2 \  , \label{gen-lapse-function}
\end{equation} 
where the parameter $\gamma$ takes the values $\gamma=0,\pm1 $ depending on the topology of the event horizon. The second, is an equation for the function $P(x,y)$:
\begin{equation} \label{EqP}
   \Delta_F P +  2\gamma \, e^{P(x,y)} +K (\vec{\nabla} \alpha )^2=0 \ ,
\end{equation}
where $\Delta_F= \partial_x^2+\partial_y^2$ is the flat-space Laplacian. The above is a Liouville equation with a smooth source, which can be explicitly solved according to the values of $\gamma$, as will see in the next section.

In absence of matter, when $\alpha=\Phi=0$, we recover the Schwarzschild-de Sitter (or anti-de Sitter) black hole with a spherical, toroidal or hyperbolic angular section. In each case, the geometry, call it $\Sigma_\gamma$, is described in conformal coordinates by
\begin{equation}\label{bm}
    e^{P_\gamma}=4(1+\gamma \rho^2)^{-2} \ , \ \ \ \  \rho^2=x^2+y^2 \ ,
\end{equation}
for $\gamma=\pm 1$, and, for $\gamma=0$ we use the more standard  $P_0=0$. Unless $\gamma=0$, we will assume $\Sigma_\gamma$ to be compact. For the hyperbolic case, that means that identifications should be performed in the $(x,y)$-space.

\section{The geometry of the horizon}

For each topology characterized by \(\gamma\), it is convenient to separate the
conformal factor \(P\) into a fixed background piece encoding the reference
geometry \(\Sigma_\gamma\) and a smooth deformation \(u\), namely
\begin{equation}
P = P_\gamma + u \, .
\end{equation}
The function \(u\) can then be regarded as a globally smooth scalar field on the
compact surface \(\Sigma_\gamma\), encoding the dynamical deformation induced by the vortices.

With this decomposition, the Liouville equation can be written intrinsically on the background geometry \(\Sigma_\gamma\) as
\begin{equation}
\Delta_\gamma u
+ 2\gamma \left(e^{u}-1\right)
= -K e^{-P_\gamma} (\vec{\nabla} \alpha )^2 \, ,
\label{liou2}
\end{equation}
where \(\Delta_\gamma = e^{-P_\gamma}\Delta_F\) is the Laplace operator
associated with the metric on $\Sigma_\gamma$.

\subsection{Area of the horizon}

To find the allowed solutions of the Liouville equation, we must demand that the area of any surface of constant $r$ must be positive.  Integrating Eq. \eqref{liou2} over the $\Sigma_\gamma$,
we obtain 
\begin{equation}\label{restr}
    A=  A_\gamma - \frac{K}{2\gamma} \int d^2 x (\vec{\nabla} \alpha )^2 > 0 \ ,
\end{equation}
where $A$, $A_\Sigma$ are the area of $\Sigma$ and $\Sigma_\gamma$ respectively.
In the hyperbolic case, it is always the case that $A>0$. In the flat case is possible only in the vacuum for compact horizons. In the spherical case, Eq. \eqref{restr} is a bound on the integral, which we will explore in detail below.

The area of the horizon is $A\, r_H^2$, where $r_H$ is the radius of the horizon; $f(r_H)=0$. Let us now compute the integral appearing in the above formulas. Using the BPS equations, one finds
\begin{equation*}
\int d^2 x (\vec{\nabla} \alpha )^2 =\int d^2 x \sin\alpha\,
(\partial_x\alpha\,\partial_y\Phi-\partial_y\alpha\,\partial_x\Phi) \ ,
\end{equation*}
which is proportional to the topological term in Eq. \eqref{eq:EBPS}, and it can be easily computed for the explicit solution in Eqs. \eqref{Hmulti}, \eqref{Phimulti} above.

Note that $\sin\alpha \, d\alpha\wedge d\Phi= -d(\cos\alpha  d\Phi)$.
This holds on the whole manifold $\Sigma$ once small disks around the singular points $\vec\rho_i$, where the vortices are located, are removed. This is because near each of these,
$\Phi \sim -\,q_i\,\varphi_i$,
where $\varphi_i$ is the polar angle around $\vec\rho_i$. 
Integrating Eq. \eqref{eq:EBPS} over the sphere with the singular regions removed, one finds
\begin{equation}
Q=-\frac{1}{4\pi}\sum_i \cos\alpha(\vec\rho_i)\oint_{\gamma_i} d\Phi =\frac{1}{2}\sum_i |q_i| \ ,
\label{eq:boundary}
\end{equation}
where $\gamma_i$ is the boundary of the small removed disk   around $\vec\rho_i$.  In deriving this we have used that, from Eqs. \eqref{alpha} and \eqref{Hmulti}, on each vortex 
$\alpha = 0$ for $q_i>0$, while $\alpha =\pi$ for $q_i<0$.
Therefore,
\begin{equation}\label{int}
    \int d^2 x (\vec{\nabla} \alpha )^2= 2\pi \sum_i |q_i| \ .
\end{equation}
The complete analysis of dynamical stability is beyond the scope of this manuscript. Nevertheless, the existence of a quantized topological charge, which appears explicitly through a discrete parameter in the solutions, is a highly desirable feature (absent in most of the available examples in the literature) that suggests intrinsic robustness and relevance of these solutions. We will come back on this issue in a future publication.

\subsection{Effect of the vortices on the horizon}

In general, the Liouville equation with sources \eqref{liou2} can only  be integrated numerically. At the horizon $r=r_H$, this describes the deformation from spherical symmetry, the $u$ function, that the horizon gets due to the presence of matter through $H$ and $\Phi$. However, some of its behavior may be deduced qualitatively. 

Although the function $H$ diverges at the vortex cores, the source term
$(\vec\nabla\alpha)^2$ entering the horizon equation remains smooth.
For a vortex of vorticity $q$, one finds, at a distance $s$ very near the core
\begin{equation}
(\vec\nabla \alpha)^2 \sim 4q^2\,s^{2|q|-2} \ ,
\end{equation}
so that for $|q|=1$ the curvature peaks at the center, while for $|q|>1$ it is
suppressed at the core and concentrates on a narrow ring around it.

Consider now a pair of vortices with opposite vorticity $\pm q$ separated by a distance $\ell$. The curvature in this case peaks in between them.
As $\ell\to0$, $H$ vanishes, but their derivatives between the vortices increase without limit. The source
$(\nabla\alpha)^2$ becomes localized in an arbitrarily small neighborhood of the coincidence point, while its integral in Eq.  \eqref{int} remains fixed by the topological charge. Geometrically, using $R_\Sigma\sqrt{g_\Sigma}=-\Delta_F P$, and integrating Eq. \eqref{EqP} on a shrinking disk,
the regular Liouville term $2\gamma e^P$ gives no contribution, and one finds in the collapsed limit
\begin{equation}
R_\Sigma(\vec\rho)\;\longrightarrow\;4\pi K|q|\,\frac{\delta^{(2)}(\vec\rho)}{\sqrt{g_{\Sigma}}} \ .
\end{equation}
This corresponds to a localized conical defect on the horizon generated
by the collapsing vortex pair, and produces a spike on it: a conical defect with angular deficit $2\pi K|q|$. We may sprinkle these defects on the horizon to produce spiky black holes. The conical defects extend throughout spacetime as cosmic strings impinging on the horizon and reaching infinity.

We now begin the analysis starting with the case $\gamma=1$ (when the cosmological constant is not necessary to have an event horizon). 

\subsection{Topologically spherical black holes}

In the spherical case, $A_{\gamma=1}=4\pi$, which may be substituted, along with Eq. \eqref{int} in Eq. \eqref{restr}, to give
\begin{equation}\label{rest1}
    \frac{K}{4} \sum_i |q_i| <1 \ ,
\end{equation}
which restricts the amount of vortices one may add on the sphere.

A simple analytic solution can be found for axially symmetric configurations 
\begin{eqnarray}
    H(\vec \rho) &=& \log\left(\rho\right) \ , \quad   \Phi(\vec \rho) = - \phi \ ,
    \label{phipair}
\end{eqnarray}
which corresponds to a pair of vortices. One with $q=1$ in the north pole, the other with $q=-1$ in the south. Naively, Eq. \eqref{phipair} does not seem to have the southern vortex, which would contradict the vanishing vorticity condition in Eq. \eqref{qz} for a compact horizon. However, the south pole is not covered by the stereographic chart centered on the north pole. To see this, one performs a $\pi$ rotation of the sphere $z\rightarrow 1/z$ in Eq. \eqref{complex}, inverting the poles.  We obtain a new solution, where the anti-vortex $q=-1$ that was hidden in the south pole, manifest itself at $\rho=0$.
 
Substituting Eq. \eqref{phipair} in Eq. \eqref{liou2} one can see that there exists a constant solution of the Liouville equation, $u=\log (1-K/2)$, hence 
\begin{equation}
e^{P}=\left(1-\frac{K}{2}\right)\frac{4}{(1+\rho^2)^2} \ .
\end{equation}
Despite the pair of vortices break the rotational symmetry in the matter sector, it is recovered for the gravitational field. $\Sigma$ is, again, a sphere, but with a radius reduced by the factor $(1-K/2)$, which is assumed to be positive because of Eq. \eqref{rest1} above. The small experimental value of $K$ is consistent with this restriction.

The resulting space-time coincides with the metric of a Schwarzschild–(A)dS black hole with a global monopole (solid angle deficit), although the deficit is sourced by BPS vortices in the NLSM rather than by a scalar triplet.

The case $\gamma=-1$ is closely analogous to the
spherical construction presented above. In this case, compact horizons are obtained by performing
discrete identifications of the Poincar\'e disk, leading to hyperbolic surfaces of genus $g>1$
(see, for instance, Refs. \cite{brill,Vanzo:1997gw}). The reduced symmetry of the resulting geometries
makes the search for fully analytic solutions more involved. Nevertheless, Eq. \eqref{restr} does not impose any restriction on the number of vortices, allowing for arbitrary
vorticity configurations on higher--genus horizons.

\subsection{Mass and Entropy}

 The mass can be derived using the ADM formula \cite{Arnowitt:1962hi}. Because the metric has the exact same form than Schwarzschild or its dS or AdS counterparts, the only difference resides in the integral over the $\Sigma$ space, namely, the area $A$ in Eq. \eqref{restr}. In the $\gamma=1$ case, for instance, the mass is
 \begin{equation}
     M = \biggl(1-\frac{K}{4} \sum_{i=1} |q_i| \biggl) m \ .
 \end{equation}
The entropy is also straightforward to obtain \cite{Wald:1993nt}, 
\begin{equation} 
S = 4 \pi \biggl(1-\frac{K}{4} \sum_{i=1} |q_i| \biggl) m^2 \ .
\end{equation}
The above expressions are typical of configurations with angular defects like global monopoles \cite{Barriola:1989hx}.

\subsection{Non-compact flat horizons}

Black objects with vortices and flat horizons cannot exist if the horizon geometry is compact. We will permit, therefore, for this case, a non-compact geometry, which represents black strings or branes. These solutions are fundamental in the holographic description of the entropy production and viscosity bounds related to broken translational symmetry \cite{Hartnoll:2016tri}. In order to have an event horizon, we see from Eq. \eqref{gen-lapse-function} that the cosmological constant must be negative.  Now Eq. \eqref{liou2}, that describes the deviation $u$ from the flat two-dimensional plane $\Sigma_0$, corresponds to the Laplace equation with a source,
\begin{eqnarray}\label{laplace}
    \Delta u =-\frac{K}{4} \left(\vec{\nabla}\alpha \right)^2.
\end{eqnarray}
We will assume that the horizon is topologically a plane where a finite quantity of vortices have been sprinkled. Hence, far away from them, for $\rho\rightarrow\infty$, we expect the geometry to be locally flat, with a possible angle deficit at infinity. In this situation, the topological charge $Q$ receives a contribution from infinity,
\begin{equation}
    Q= \frac{1}{2}\sum_i |q_i| + \frac{1}{2}\left|\sum_i q_i\right|,
\end{equation}
and it turns out that the deficit angle of the plane at infinity is $\pi K Q$. The fact that this deficit must be less than $2\pi$ implies the restriction $KQ/2< 1$ in the number of vortices, analog to Eq. \eqref{rest1} in the spherical case.

When every $q_{i}$ has the same sign (say positive) Eq. \eqref{laplace} has an explicit solution which gives the following conformal factor for the metric of $\Sigma$:
\begin{equation}\label{flatsol}
    e^P=\left(1+\prod_{i}|\vec{\rho}-\vec{\rho}_{i}|^{2q_{i}} \right)^{-K} \ . 
\end{equation}
When all vorticities have the same sign, Eq. \eqref{flatsol} yields an explicit analytic black brane solution whose horizon geometry is fully controlled by the quantized superfluid vortices. Also in this case, the number of bumps in the curvature of the horizon is related to the vorticity of the pions. An interesting question (on which we hope to come back in the future) is whether one can choose the positions of the bumps in such a way to have a periodic solution and a compact horizon.

\section{Final remarks}

We have constructed the first analytic examples of black holes and black branes with bumpy horizons supported by realistic matter fields (pions) in GR minimally coupled with the NLSM. The number of bumps of the horizon is related to the quantized vorticity of the pions field, which exactly satisfies the Onsager-Feynman relation for the quantum superfluid phase. This discrete topological charge enters directly in the first law of black hole thermodynamics. Furthermore, our results allow an explicit realization of the holographic description of the entropy production and viscosity bounds due to broken translational symmetry.

While astrophysical black holes likely involve diverse matter contributions, the role of pionic matter is uniquely significant due to its topological robustness. We have demonstrated that the resulting horizon bumps are associated with BPS topological vortices; since vorticity is topologically conserved, these bumps cannot be continuously turned off. This topological protection ensures that the signature of pionic superfluidity remains stable even when additional matter fields are present. Consequently, despite the complexities of realistic astrophysical environments, pionic vortices will manifest as persistent deformations of the black hole horizon. 

An interesting outcome of the present paper is the following: bumps on the horizons are a fingerprint of superfluid pionic vortices. Such a fingerprint is topologically protected as vorticity is topologically conserved. Therefore, it is very interesting to search for such a fingerprint in astrophysical relevant situations where superfluid pionic vortices play a fundamental role (see, for instance, \cite{Poli:2023vyp} and references therein). We will return to this issue in a future publication.

\begin{acknowledgments}
F.C. has been supported by FONDECYT Grant No. 1240048 and by Grant ANID EXPLORACION No. 13250014. C. H. appreciates the support of FONDECYT postdoctoral Grant No. 3240632. A. V.
has been funded by FONDECYT Iniciaci\'on No. 11261883.
\end{acknowledgments}

\bibliography{apssamp}

@PREAMBLE{
 "\providecommand{\noopsort}[1]{}" 
 # "\providecommand{\singleletter}[1]{#1}%" 
}

@article{Israel:1967wq,
    author = "Israel, Werner",
    title = "{Event horizons in static vacuum space-times}",
    doi = "10.1103/PhysRev.164.1776",
    journal = "Phys. Rev.",
    volume = "164",
    pages = "1776--1779",
    year = "1967"
}

@article{Carter:1971zc,
    author = "Carter, B.",
    title = "{Axisymmetric Black Hole Has Only Two Degrees of Freedom}",
    doi = "10.1103/PhysRevLett.26.331",
    journal = "Phys. Rev. Lett.",
    volume = "26",
    pages = "331--333",
    year = "1971"
}

@article{Hawking:1971vc,
    author = "Hawking, S. W.",
    title = "{Black holes in general relativity}",
    doi = "10.1007/BF01877517",
    journal = "Commun. Math. Phys.",
    volume = "25",
    pages = "152--166",
    year = "1972"
}

@article{Robinson:1975bv,
    author = "Robinson, D. C.",
    title = "{Uniqueness of the Kerr black hole}",
    doi = "10.1103/PhysRevLett.34.905",
    journal = "Phys. Rev. Lett.",
    volume = "34",
    pages = "905--906",
    year = "1975"
}

@article{Friedman:1993ty,
    author = "Friedman, John L. and Schleich, Kristin and Witt, Donald M.",
    title = "{Topological censorship}",
    eprint = "gr-qc/9305017",
    archivePrefix = "arXiv",
    reportNumber = "PRINT-93-0448 (SANTA-BARBARA,ITP), NSF-ITP-93-80",
    doi = "10.1103/PhysRevLett.71.1486",
    journal = "Phys. Rev. Lett.",
    volume = "71",
    pages = "1486--1489",
    year = "1993",
    note = "[Erratum: Phys.Rev.Lett. 75, 1872 (1995)]"
}

@article{Gibbons:1977mu,
    author = "Gibbons, G. W. and Hawking, S. W.",
    title = "{Cosmological Event Horizons, Thermodynamics, and Particle Creation}",
    doi = "10.1103/PhysRevD.15.2738",
    journal = "Phys. Rev. D",
    volume = "15",
    pages = "2738--2751",
    year = "1977"
}

@article{Linet:1986sr,
    author = "Linet, B.",
    title = "{The static, cylindrically symmetric strings in general relativity with cosmological constant}",
    doi = "10.1063/1.527050",
    journal = "J. Math. Phys.",
    volume = "27",
    pages = "1817--1818",
    year = "1986"
}

@article{Lemos:1994xp,
    author = "Lemos, J. P. S.",
    title = "{Cylindrical black hole in general relativity}",
    eprint = "gr-qc/9404041",
    archivePrefix = "arXiv",
    doi = "10.1016/0370-2693(95)00533-Q",
    journal = "Phys. Lett. B",
    volume = "353",
    pages = "46--51",
    year = "1995"
}

@article{Birmingham:1998nr,
    author = "Birmingham, Danny",
    title = "{Topological black holes in Anti-de Sitter space}",
    eprint = "hep-th/9808032",
    archivePrefix = "arXiv",
    doi = "10.1088/0264-9381/16/4/009",
    journal = "Class. Quant. Grav.",
    volume = "16",
    pages = "1197--1205",
    year = "1999"
}

@article{Klemm:2014rda,
    author = "Klemm, Dietmar",
    title = "{Four-dimensional black holes with unusual horizons}",
    eprint = "1401.3107",
    archivePrefix = "arXiv",
    primaryClass = "hep-th",
    reportNumber = "IFUM-1022-FT",
    doi = "10.1103/PhysRevD.89.084007",
    journal = "Phys. Rev. D",
    volume = "89",
    number = "8",
    pages = "084007",
    year = "2014"
}

@article{Cai:1996eg,
    author = "Cai, Rong-Gen and Zhang, Yuan-Zhong",
    title = "{Black plane solutions in four-dimensional space-times}",
    eprint = "gr-qc/9609065",
    archivePrefix = "arXiv",
    doi = "10.1103/PhysRevD.54.4891",
    journal = "Phys. Rev. D",
    volume = "54",
    pages = "4891--4898",
    year = "1996"
}

@article{Vanzo:1997gw,
    author = "Vanzo, Luciano",
    title = "{Black holes with unusual topology}",
    eprint = "gr-qc/9705004",
    archivePrefix = "arXiv",
    reportNumber = "UTF-400",
    doi = "10.1103/PhysRevD.56.6475",
    journal = "Phys. Rev. D",
    volume = "56",
    pages = "6475--6483",
    year = "1997"
}

@article{Emparan:2001wn,
    author = "Emparan, Roberto and Reall, Harvey S.",
    title = "{A Rotating black ring solution in five-dimensions}",
    eprint = "hep-th/0110260",
    archivePrefix = "arXiv",
    reportNumber = "CERN-TH-2001-294",
    doi = "10.1103/PhysRevLett.88.101101",
    journal = "Phys. Rev. Lett.",
    volume = "88",
    pages = "101101",
    year = "2002"
}

@article{Elvang:2007rd,
    author = "Elvang, Henriette and Figueras, Pau",
    title = "{Black Saturn}",
    eprint = "hep-th/0701035",
    archivePrefix = "arXiv",
    reportNumber = "MIT-CTP-3796",
    doi = "10.1088/1126-6708/2007/05/050",
    journal = "JHEP",
    volume = "05",
    pages = "050",
    year = "2007"
}

@article{Gubser:2001ac,
    author = "Gubser, Steven S.",
    title = "{On nonuniform black branes}",
    eprint = "hep-th/0110193",
    archivePrefix = "arXiv",
    reportNumber = "CALT-68-2351, CITUSC-01-035",
    doi = "10.1088/0264-9381/19/19/303",
    journal = "Class. Quant. Grav.",
    volume = "19",
    pages = "4825--4844",
    year = "2002"
}

@article{Licht:2020odx,
    author = "Licht, David and Luna, Raimon and Suzuki, Ryotaku",
    title = "{Black Ripples, Flowers and Dumbbells at large $D$}",
    eprint = "2002.07813",
    archivePrefix = "arXiv",
    primaryClass = "hep-th",
    doi = "10.1007/JHEP04(2020)108",
    journal = "JHEP",
    volume = "04",
    pages = "108",
    year = "2020"
}

@article{Iguchi:2007is,
    author = "Iguchi, Hideo and Mishima, Takashi",
    title = "{Black di-ring and infinite nonuniqueness}",
    eprint = "hep-th/0701043",
    archivePrefix = "arXiv",
    doi = "10.1103/PhysRevD.78.069903",
    journal = "Phys. Rev. D",
    volume = "75",
    pages = "064018",
    year = "2007",
    note = "[Erratum: Phys.Rev.D 78, 069903 (2008)]"
}

@article{Chen:2016rjt,
    author = "Chen, Yu and Teo, Edward",
    title = "{Black holes with bottle-shaped horizons}",
    eprint = "1604.07527",
    archivePrefix = "arXiv",
    primaryClass = "gr-qc",
    doi = "10.1103/PhysRevD.93.124028",
    journal = "Phys. Rev. D",
    volume = "93",
    number = "12",
    pages = "124028",
    year = "2016"
}

@article{Ferrero:2020twa,
    author = "Ferrero, Pietro and Gauntlett, Jerome P. and Ipi{\~n}a, Juan Manuel P{\'e}rez and Martelli, Dario and Sparks, James",
    title = "{Accelerating black holes and spinning spindles}",
    eprint = "2012.08530",
    archivePrefix = "arXiv",
    primaryClass = "hep-th",
    doi = "10.1103/PhysRevD.104.046007",
    journal = "Phys. Rev. D",
    volume = "104",
    number = "4",
    pages = "046007",
    year = "2021"
}

@article{Ferrero:2020laf,
    author = "Ferrero, Pietro and Gauntlett, Jerome P. and P{\'e}rez Ipi{\~n}a, Juan Manuel and Martelli, Dario and Sparks, James",
    title = "{D3-Branes Wrapped on a Spindle}",
    eprint = "2011.10579",
    archivePrefix = "arXiv",
    primaryClass = "hep-th",
    doi = "10.1103/PhysRevLett.126.111601",
    journal = "Phys. Rev. Lett.",
    volume = "126",
    number = "11",
    pages = "111601",
    year = "2021"
}

@article{Giri:2021xta,
    author = "Giri, Suvendu",
    title = "{Black holes with spindles at the horizon}",
    eprint = "2112.04431",
    archivePrefix = "arXiv",
    primaryClass = "hep-th",
    doi = "10.1007/JHEP06(2022)145",
    journal = "JHEP",
    volume = "06",
    pages = "145",
    year = "2022"
}

@article{Boido:2022mbe,
    author = "Boido, Andrea and Gauntlett, Jerome P. and Martelli, Dario and Sparks, James",
    title = "{Gravitational Blocks, Spindles and GK Geometry}",
    eprint = "2211.02662",
    archivePrefix = "arXiv",
    primaryClass = "hep-th",
    reportNumber = "Imperial/TP/2022/JG/04",
    doi = "10.1007/s00220-023-04812-8",
    journal = "Commun. Math. Phys.",
    volume = "403",
    number = "2",
    pages = "917--1003",
    year = "2023"
}

@article{Novikov:1992gow,
    author = "Novikov, I. D. and Manko, V. S.",
    title = "{Generalizations of the Kerr and Kerr-Newman metrics possessing an arbitrary set of mass-multipole moments}",
    doi = "10.1088/0264-9381/9/11/013",
    journal = "Class. Quant. Grav.",
    volume = "9",
    number = "11",
    pages = "2477",
    year = "1992"
}

@article{Collins:2004ex,
    author = "Collins, Nathan A. and Hughes, Scott A.",
    title = "{Towards a formalism for mapping the space-times of massive compact objects: Bumpy black holes and their orbits}",
    eprint = "gr-qc/0402063",
    archivePrefix = "arXiv",
    doi = "10.1103/PhysRevD.69.124022",
    journal = "Phys. Rev. D",
    volume = "69",
    pages = "124022",
    year = "2004"
}

@article{Johannsen:2011dh,
    author = "Johannsen, Tim and Psaltis, Dimitrios",
    title = "{A Metric for Rapidly Spinning Black Holes Suitable for Strong-Field Tests of the No-Hair Theorem}",
    eprint = "1105.3191",
    archivePrefix = "arXiv",
    primaryClass = "gr-qc",
    doi = "10.1103/PhysRevD.83.124015",
    journal = "Phys. Rev. D",
    volume = "83",
    pages = "124015",
    year = "2011"
}

@article{Vigeland:2011ji,
    author = "Vigeland, Sarah and Yunes, Nicolas and Stein, Leo",
    title = "{Bumpy Black Holes in Alternate Theories of Gravity}",
    eprint = "1102.3706",
    archivePrefix = "arXiv",
    primaryClass = "gr-qc",
    doi = "10.1103/PhysRevD.83.104027",
    journal = "Phys. Rev. D",
    volume = "83",
    pages = "104027",
    year = "2011"
}

@article{Emparan:2014pra,
    author = "Emparan, Roberto and Figueras, Pau and Martinez, Marina",
    title = "{Bumpy black holes}",
    eprint = "1410.4764",
    archivePrefix = "arXiv",
    primaryClass = "hep-th",
    doi = "10.1007/JHEP12(2014)072",
    journal = "JHEP",
    volume = "12",
    pages = "072",
    year = "2014"
}

@article{Dubovsky:2007zi,
    author = "Dubovsky, Sergei and Tinyakov, Peter and Zaldarriaga, Matias",
    title = "{Bumpy black holes from spontaneous Lorentz violation}",
    eprint = "0706.0288",
    archivePrefix = "arXiv",
    primaryClass = "hep-th",
    doi = "10.1088/1126-6708/2007/11/083",
    journal = "JHEP",
    volume = "11",
    pages = "083",
    year = "2007"
}

@article{Hristov:2023rel,
    author = "Hristov, Kiril and Suh, Minwoo",
    title = "{Spindle black holes in AdS$_{4} \times$ SE$_{7}$}",
    eprint = "2307.10378",
    archivePrefix = "arXiv",
    primaryClass = "hep-th",
    doi = "10.1007/JHEP10(2023)141",
    journal = "JHEP",
    volume = "10",
    pages = "141",
    year = "2023"
}

@article{Suh:2023xse,
    author = "Suh, Minwoo",
    title = "{Baryonic spindles from conifolds}",
    eprint = "2304.03308",
    archivePrefix = "arXiv",
    primaryClass = "hep-th",
    doi = "10.1007/JHEP02(2025)181",
    journal = "JHEP",
    volume = "02",
    pages = "181",
    year = "2025"
}

@article{Farhangkhah:2014zka,
    author = "Farhangkhah, N. and Dehghani, M. H.",
    title = "{Lovelock black holes with nonmaximally symmetric horizons}",
    eprint = "1409.1410",
    archivePrefix = "arXiv",
    primaryClass = "gr-qc",
    doi = "10.1103/PhysRevD.90.044014",
    journal = "Phys. Rev. D",
    volume = "90",
    number = "4",
    pages = "044014",
    year = "2014"
}

@article{Gregory:1993vy,
    author = "Gregory, R. and Laflamme, R.",
    title = "{Black strings and p-branes are unstable}",
    eprint = "hep-th/9301052",
    archivePrefix = "arXiv",
    doi = "10.1103/PhysRevLett.70.2837",
    journal = "Phys. Rev. Lett.",
    volume = "70",
    pages = "2837--2840",
    year = "1993"
}

@article{Lehner:2010pn,
    author = "Lehner, Luis and Pretorius, Frans",
    title = "{Black Strings, Low Viscosity Fluids, and Violation of Cosmic Censorship}",
    eprint = "1006.5960",
    archivePrefix = "arXiv",
    primaryClass = "hep-th",
    doi = "10.1103/PhysRevLett.105.101102",
    journal = "Phys. Rev. Lett.",
    volume = "105",
    pages = "101102",
    year = "2010"
}

@article{Figueras:2015hkb,
    author = "Figueras, Pau and Kunesch, Markus and Tunyasuvunakool, Saran",
    title = "{End Point of Black Ring Instabilities and the Weak Cosmic Censorship Conjecture}",
    eprint = "1512.04532",
    archivePrefix = "arXiv",
    primaryClass = "hep-th",
    doi = "10.1103/PhysRevLett.116.071102",
    journal = "Phys. Rev. Lett.",
    volume = "116",
    number = "7",
    pages = "071102",
    year = "2016"
}

@article{Herdeiro:2014goa,
    author = "Herdeiro, Carlos A. R. and Radu, Eugen",
    title = "{Kerr black holes with scalar hair}",
    eprint = "1403.2757",
    archivePrefix = "arXiv",
    primaryClass = "gr-qc",
    doi = "10.1103/PhysRevLett.112.221101",
    journal = "Phys. Rev. Lett.",
    volume = "112",
    pages = "221101",
    year = "2014"
}

@article{Herdeiro:2016tmi,
    author = "Herdeiro, Carlos and Radu, Eugen and R{\'u}narsson, Helgi",
    title = "{Kerr black holes with Proca hair}",
    eprint = "1603.02687",
    archivePrefix = "arXiv",
    primaryClass = "gr-qc",
    doi = "10.1088/0264-9381/33/15/154001",
    journal = "Class. Quant. Grav.",
    volume = "33",
    number = "15",
    pages = "154001",
    year = "2016"
}

@article{Doneva:2017bvd,
    author = "Doneva, Daniela D. and Yazadjiev, Stoytcho S.",
    title = "{New Gauss-Bonnet Black Holes with Curvature-Induced Scalarization in Extended Scalar-Tensor Theories}",
    eprint = "1711.01187",
    archivePrefix = "arXiv",
    primaryClass = "gr-qc",
    doi = "10.1103/PhysRevLett.120.131103",
    journal = "Phys. Rev. Lett.",
    volume = "120",
    number = "13",
    pages = "131103",
    year = "2018"
}

@article{Silva:2017uqg,
    author = "Silva, Hector O. and Sakstein, Jeremy and Gualtieri, Leonardo and Sotiriou, Thomas P. and Berti, Emanuele",
    title = "{Spontaneous scalarization of black holes and compact stars from a Gauss-Bonnet coupling}",
    eprint = "1711.02080",
    archivePrefix = "arXiv",
    primaryClass = "gr-qc",
    doi = "10.1103/PhysRevLett.120.131104",
    journal = "Phys. Rev. Lett.",
    volume = "120",
    number = "13",
    pages = "131104",
    year = "2018"
}

@article{Doneva:2022ewd,
    author = "Doneva, Daniela D. and Ramazano{\u{g}}lu, Fethi M. and Silva, Hector O. and Sotiriou, Thomas P. and Yazadjiev, Stoytcho S.",
    title = "{Spontaneous scalarization}",
    eprint = "2211.01766",
    archivePrefix = "arXiv",
    primaryClass = "gr-qc",
    doi = "10.1103/RevModPhys.96.015004",
    journal = "Rev. Mod. Phys.",
    volume = "96",
    number = "1",
    pages = "015004",
    year = "2024"
}

@article{onsager1949,
  title={Statistical hydrodynamics},
  author={Onsager, Lars},
  journal={Il Nuovo Cimento},
  volume={6},
  number={Supplement 2},
  pages={279--287},
  year={1949}
}

@incollection{feynman1955,
  title={Application of quantum mechanics to liquid helium},
  author={Feynman, Richard P},
  booktitle={Progress in Low Temperature Physics},
  volume={1},
  pages={17--53},
  year={1955},
  publisher={Elsevier}
}

@book{donnelly1991,
  title={Quantized Vortices in Helium II},
  author={Donnelly, Russell J.},
  year={1991},
  publisher={Cambridge University Press},
  series={Cambridge Studies in Low Temperature Physics},
  volume={3},
  address={Cambridge},
  isbn={9780521324007}
}

@article{fetter2009,
  title = {Rotating trapped Bose-Einstein condensates},
  author = {Fetter, Alexander L.},
  journal = {Reviews of Modern Physics},
  volume = {81},
  issue = {2},
  pages = {647--691},
  year = {2009},
  month = {Apr},
  publisher = {American Physical Society},
  doi = {10.1103/RevModPhys.81.647},
  url = {https://link.aps.org/doi/10.1103/RevModPhys.81.647}
}

@article{Canfora:2024mkp,
    author = "Canfora, Fabrizio and Lagos, Marcela and Vera, Aldo",
    title = "{Superconducting multi-vortices and a novel BPS bound in chiral perturbation theory}",
    eprint = "2405.08082",
    archivePrefix = "arXiv",
    primaryClass = "hep-th",
    doi = "10.1007/JHEP10(2024)224",
    journal = "JHEP",
    volume = "10",
    pages = "224",
    year = "2024"
}

@article{Hartnoll:2016tri,
    author = "Hartnoll, Sean A. and Ramirez, David M. and Santos, Jorge E.",
    title = "{Entropy production, viscosity bounds and bumpy black holes}",
    eprint = "1601.02757",
    archivePrefix = "arXiv",
    primaryClass = "hep-th",
    doi = "10.1007/JHEP03(2016)170",
    journal = "JHEP",
    volume = "03",
    pages = "170",
    year = "2016"
}

@article{Kovtun:2004de,
    author = "Kovtun, P. and Son, Dan T. and Starinets, Andrei O.",
    title = "{Viscosity in strongly interacting quantum field theories from black hole physics}",
    eprint = "hep-th/0405231",
    archivePrefix = "arXiv",
    reportNumber = "INT-PUB-04-09, UW-PT-04-04",
    doi = "10.1103/PhysRevLett.94.111601",
    journal = "Phys. Rev. Lett.",
    volume = "94",
    pages = "111601",
    year = "2005"
}

@article{Poli:2023vyp,
    author = "Poli, Elena and Bland, Thomas and White, Samuel J. M. and Mark, Manfred J. and Ferlaino, Francesca and Trabucco, Silvia and Mannarelli, Massimo",
    title = "{Glitches in Rotating Supersolids}",
    eprint = "2306.09698",
    archivePrefix = "arXiv",
    primaryClass = "cond-mat.quant-gas",
    doi = "10.1103/PhysRevLett.131.223401",
    journal = "Phys. Rev. Lett.",
    volume = "131",
    number = "22",
    pages = "223401",
    year = "2023"
}

@article{brill,
  title = {Thermodynamics of $(3+1)$-dimensional black holes with toroidal or higher genus horizons},
  author = {Brill, Dieter R. and Louko, Jorma and Peld\'an, Peter},
  journal = {Phys. Rev. D},
  volume = {56},
  issue = {6},
  pages = {3600--3610},
  numpages = {0},
  year = {1997},
  month = {Sep},
  publisher = {American Physical Society},
  doi = {10.1103/PhysRevD.56.3600},
  url = {https://link.aps.org/doi/10.1103/PhysRevD.56.3600}
}

@article{Arnowitt:1962hi,
    author = "Arnowitt, Richard L. and Deser, Stanley and Misner, Charles W.",
    title = "{The Dynamics of general relativity}",
    eprint = "gr-qc/0405109",
    archivePrefix = "arXiv",
    doi = "10.1007/s10714-008-0661-1",
    journal = "Gen. Rel. Grav.",
    volume = "40",
    pages = "1997--2027",
    year = "2008"
}

@article{Wald:1993nt,
    author = "Wald, Robert M.",
    title = "{Black hole entropy is the Noether charge}",
    eprint = "gr-qc/9307038",
    archivePrefix = "arXiv",
    reportNumber = "EFI-93-42",
    doi = "10.1103/PhysRevD.48.R3427",
    journal = "Phys. Rev. D",
    volume = "48",
    number = "8",
    pages = "R3427--R3431",
    year = "1993"
}

@article{Barriola:1989hx,
    author = "Barriola, Manuel and Vilenkin, Alexander",
    title = "{Gravitational Field of a Global Monopole}",
    reportNumber = "TUTP-89-4",
    doi = "10.1103/PhysRevLett.63.341",
    journal = "Phys. Rev. Lett.",
    volume = "63",
    pages = "341",
    year = "1989"
}

\end{document}